\documentstyle[12pt,aaspp4]{article}

\begin{document}

\title{GRB990123, The Optical Flash and The Fireball Model}
\author{Re'em Sari$^1$ and Tsvi Piran$^2$ \\
$^1$ Theoretical Astrophysics 130-33, California Institute of Technology,
Pasadena, CA 91125, USA \\
$^2$ Racah Institute of Physics, The Hebrew University, Jerusalem 91904, 
Israel and Department of Physics, Columbia University, New York, NY 10027, USA}

\begin{abstract}
We compare the ongoing observations of the remarkable burst GRB990123
with the predictions of the afterglow theory. 
We show that the observations agree with the recent
prediction that a reverse shock propagating into the ejecta would
produce a very strong prompt optical flash.  This reverse shock has
also produced the 8.46GHz radio signal, observed after one day. The
forward shock, which propagates into the ISM is the origin of the classical
afterglow. It  has produced the prompt X-ray signal as well as the late
optical and IR emission.  It would most likely produce a radio
emission within the next few weeks. The observations suggest that the
initial Lorentz factor of the ejecta was $\sim 200$. 
Within  factors of order unity, this crude model explains all
current observations of GRB990123.
\end{abstract}


\section{Introduction}

Gamma-ray bursts observers, were shocked once more with the explosion
of GRB990123. This is a very strong burst. Its
fluence of $3\times 10^{-4}{\rm erg/cm^2}$ (Kippen et al., GCN224) places it
at the top 0.3\% of BATSE's bursts.  It has a multi-wavelength afterglow
ranging from X-ray via optical and IR to radio. Absorption lines in
the optical have led to a lower limit of its redshift $z>1.6$ 
(Kelson et al., IAUC 7096) which
for isotropic emission leads to a $\gamma$-ray energy of about
$3\times10^{54}$ergs. This, and a second set of absorption lines at $z
\sim 0.2$ have led to the suggestion (Djorgovski et al., GCN216) that 
GRB990123 might have been lensed and amplified by a factor of ten or
so.

GRB990123 would have been amongst the most exciting GRBs ever just on
the basis of these facts.  However, ROTSE discovered a new element
8.9 magnitude prompt optical flash  (Akerlof
et al., GCN205). This have added another dimension to GRB
astronomy. It is the first time that a prompt emission in another
wavelength apart from $\gamma$-rays has been detected from a GRB.
Such a strong optical flash was predicted, just a few weeks ago (Sari
\& Piran 1999a,b; hereafter SP99), to arise from a reverse shock, propagating
into the relativistic ejecta, that forms in the early afterglow.  The
original prediction gave a lower limit of 15 magnitude for a
``standard'' GRB with a fluence of $10^{-5}$ergs/cm$^2$.  Scaling that
to the $\gamma$-ray fluence of GRB990123 yields a lower limit of $\sim
11$, compatible with the observed 9 magnitude.

In this letter we confront the fireball theory with the ongoing
observations of GRB990123.  We show that the observations of the GRB
light curve and spectrum, the prompt optical flash light curve, the
radio emission as well as the available afterglow light curve for the
first few days strongly support the reverse shock prompt
emission model. This agreement provides an additional support to the
overall internal-external scenario.

\section{Observations}

GRB990123 triggered BATSE on 1999 January 23.507594 (Kippen et al.,
GCN 224).
It consisted of multi-peaked structure lasting more than 100 seconds.
There are two clear relatively hard peaks with irregular softer
emission that follow. The burst's $\gamma$-ray peak flux is
16.42 photons/cm$^2$/sec. The total fluence ($>20$keV) is $\sim 3
\times 10^{-4}$ erg/cm$^2$ (Band 1999). 
The burst also triggered GRBM (on 23.50780) and
was detected by the WFC on BeppoSAX (Feroci et al., IAUC 7095). The GRBM
fluence is comparable $\sim 3.5 \times 10^{-4}$
erg/cm$^2$.  The WFC light curve is complex with only one clear peak
(about 40 seconds after the GRBM peak) followed by a structured high
plateau. The peak flux of this peak is $\sim 3.4$ Crab
in the energy band 1.5-26 keV. The total fluence in
this soft X-ray band is about $7 \times 10^{-6}$ergs/cm$^2$, 
a few percent of the $\gamma$-ray fluence.

BATSE's observations triggered ROTSE via the BACODINE system (Akerlof et al.,
GCN205). An 11.82 magnitude optical flash was detected on the first 5
seconds exposure, 22.18 seconds after the onset of
the burst. This was the first observation ever of a prompt optical
counterpart of a GRB.  Another 5 seconds exposure,  25
seconds later, revealed a 8.95 magnitude signal ($\sim 1$ Jy!). The optical
signal 
decayed to 10.08 magnitude 25 seconds later and  continued to
decay down to 14.53 magnitude in subsequent three 75 second exposures
that took place up to 10 minutes after the burst. The five last
exposures depict a power law decay with a slope of $\sim 2.0$ (see
Fig. 1). This initial optical flash contains most of the optical
fluence: $\sim 2.5 \times 10^{-7}{\rm ergs/cm^2}$, about $7.7 \times
10^{-4}$ of the $\gamma$-ray fluence.

\begin{figure}[tbp]
\begin{center}
\epsscale{1.} \plotone{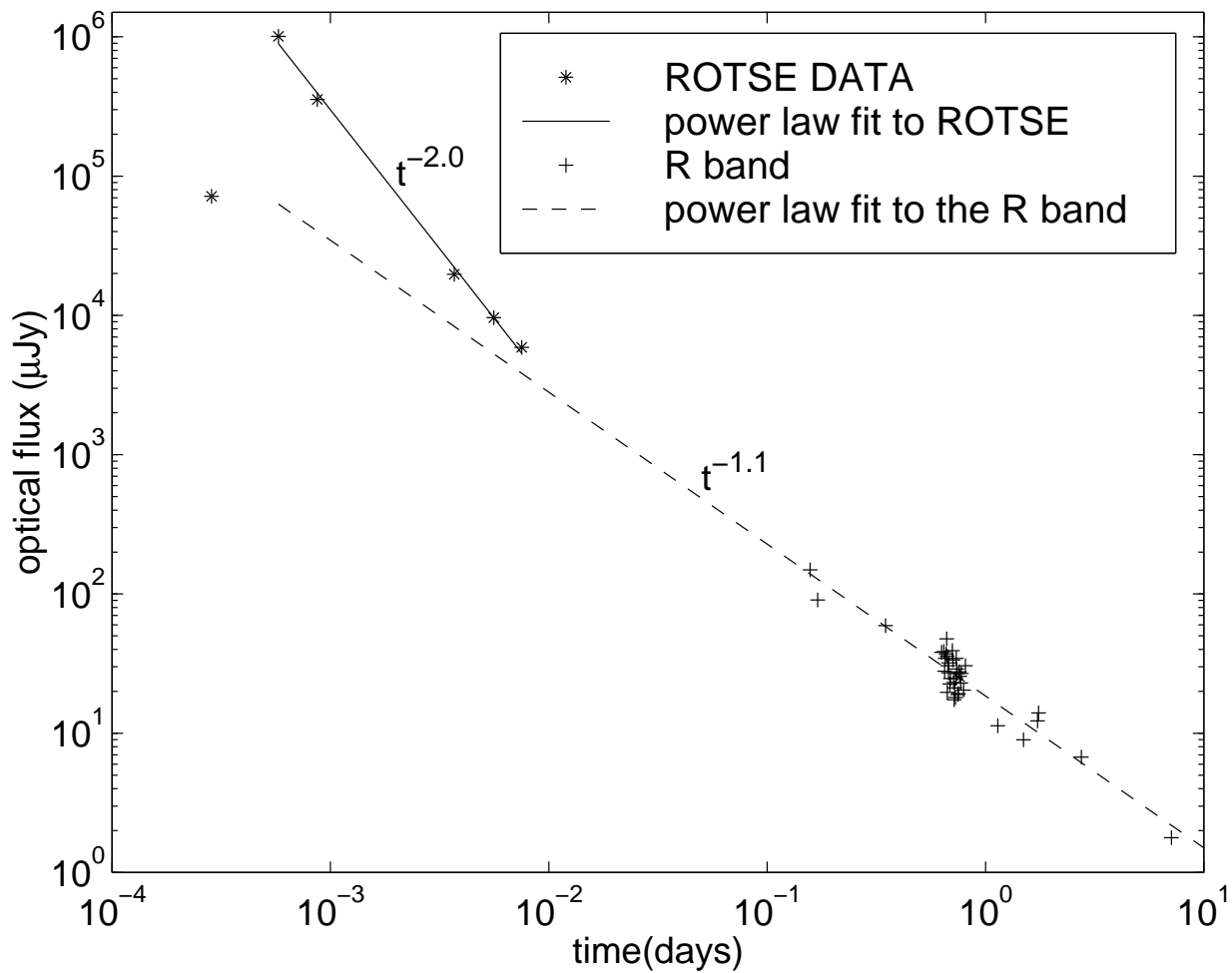}
\end{center}
\caption{The optical light curve of  GRB990123. The early ROTSE measurements, 
excluding the first one, are well described by a power law. Similarly
are the R band data points from various groups
reported in the GCN and on IAUC by the time this paper is written 
(Odewahn et al., IAUC. 7094; Gal et al., GCN207; Bloom et al., GCN208;
Zhu \& Zhang, GCN204; Sokolov et al., GCN209; Ofek \& Leibowitz,
GCN210; Masetti et al., GCN220;  Garnavich et al., GCN215; Sagar et al.,
GCN227; Yadigaroglu et al GCN242). 
The discrepancy between the two power
laws is evident.}
\label{fig1}
\end{figure}

Optical afterglow follow up by larger telescopes began some 3 hours
and 46 minutes later with the observations on Palomar (Odewahn et al.,
IAUC. 7094). These observations revealed an 18.2 R magnitude
source. The optical observations continued in more than half a dozen
observatories around the world. These observations are summarized in
Fig. 1.  We have inferred a slope of $\sim -1.1$ from the entire R band
observations reposted in the GCN (see Fig 1). 
A similar slope  $-1.13 \pm 0.03$ was deduced for the Gunn-r flux 
(Bloom et al., GCN240, Yadigaroglu et al, GCN242).
Note that this is
significantly different from the initial slope. The optical spectra
revealed several absorption line system, showing that the redshift of
GRB990123 is $\ge 1.61$ (Kelson et al., IAUC 7096).

The early X-ray observations were followed up by an NFI observation
(Piro et al., GCN203, Heise et al IAUC7099) beginning approximately six
hours after the burst with a flux of $1.1 \times 10^{-11}$
ergs/cm$^2$/sec (about $0.8 \mu$Jy) and lasting for 26hours.  The NFI
observation corresponds to a power law decay with a slope of -1.35 from
the prompt observation (about 60 seconds after the burst) and within
the 26 hour NFI measurement itself.  An ASCA observation (Murakami et
al., GCN228) on 25.688 (approximately 2 days and 7 hours after the
burst) reported a flux of $\sim 10^{-12}$ergs/cm$^2$/sec (about $0.08
\mu$Jy).  The decay from the NFI to the ASCA observation is slightly
slower with a slope of $\sim -1.1$.

A Near Infra-red counterpart with a $K=18.3 \pm 0.03$ magnitude was
detected on Jan. 24.6356 (Bloom et al., GCN240). It has been observed later
on 27.65 and 28.55.  The observations agree with a decaying light
curve with a slope $-1.14 \pm 0.08$. 

Finally, a radio source at 8.46 GHz was detected on Jan. 24.65 by the
VLA (Frail \& Kulkarni, GCN211) with a flux density of $260 \pm 32
\mu$Jy. This radio source was not detected earlier with an upper limit
of $64\mu$Jy (Frail and Kulkarni GCN200) or later with a comparable
upper limit by the VLA (Kulkarni and Frail GCN239).  Earlier attempt
to detect a radio source on 24.28 at 4.88GHz gave only an upper limit
of $130\mu$Jy (Galama et al., GCN212)

\section{An Optical Flash from the Reverse Shock}

A brief examination of the $\gamma$-rays signal during the three
optical exposures that where simultaneous with the burst show that
there is no correlation between the $\gamma$-ray intensity and the
optical intensity. The $\gamma$-ray counts ratios in these three
exposures are 5:1:1 (a more careful examination of the spectrum shows
that the energy ratio is about 10:1:1; Band, 1999). The optical
ratios, on the other hand, are approximately 1:15:5. While in
principle the cooling tail of the electrons producing the GRB itself
could give rise to a strong optical signal it would have been
correlated with the $\gamma$-ray signal.  The lack of correlations
means that the same electrons could not have emitted both the
$\gamma$-rays and the optical emission. Moreover, GRB990123 was a
relatively hard burst peaking at about $1$MeV. The decreasing flux
with decreasing frequency
(below a few hundred keV) is incompatible with a low cooling frequency,
required for a strong optical emission. One could have thought that
the highest energy electrons which are emitting $\gamma$-rays are fast
cooling while the lower energy electrons which are emitting optical
are slow cooling.  In this case the optical emission would have been
proportional to the integrated $\gamma$-ray flux (Katz, 1999). But this
again is in disagreement with the decay in the optical emission in the
third optical exposure. We conclude that the $\gamma$-rays and the
optical emission must have been emitted in different physical regions.

According to the fireball model (see e.g. Piran, 1999 for a review)
there are two possible regions in which shocks could take place.
Internal shocks which take place within the relativistic ejecta, and
external shocks that take place between the ejecta and the ISM. In the
internal-external model (Sari \& Piran 1997) the GRB is produced via
internal shocks within the relativistic ejecta itself while the
afterglow is produced via external shocks. In an internal shock both
forward and reverse are more or less similar since the ejecta shells
they are running into have similar properties. On the other hand, in
the external shocks that follow, the reverse shock, that is going
into the dense ejecta, is very different from the forward shock that
is going into the ISM.  Therefore, overall there are three possible
emitting regions in the internal-external scenario.

For external shocks the ratio between the emission from the forward
and the reverse shock is proportional to $\gamma^2$, $\gamma$ being
the Lorentz factor of the ejecta (See SP99 for more details). 
Thus, if an external reverse shock
is producing the GRB the external forward shock emission will be in
the GeV range (M\'esz\'aros and Rees, 1994), and
there is no room for a strong optical emission. If an external
forward shock has  produced the GRB the external reverse shock could
have emitted in optical.  However, we expect such emission to be
correlated with the $\gamma$-ray emission, unlike the
observations. Thus, we rule out this scenario.  This is in agreement
with other arguments against this model (Sari \& Piran 1997,
Fenimore, Madras \& Nayakshin 1996).

Within the internal-external scenario the GRB is produced by the
internal shocks.  For these shocks both forward and reverse shocks are
rather similar and the emission from both shocks is at the same energy
band. If the forward external shock would have produced the optical
emission there would have been no place to produce neither the prompt
X-rays nor the late afterglow emission. Thus, we are left with the only
possibility that the reverse external shock has produced the optical
emission while the forward external shock (which continues later as
the afterglow) produced the early X-ray as well as UV and some weak
$\gamma$-ray signal.  We don't expect now any correlation between the
$\gamma$-rays and the optical emission, but we expect some correlation
between the optical emission and an early X-ray emission. Indeed the
WFC on BeppoSAX reported an X-ray peak some 60 seconds after the
beginning of the burst, not far from the peak exposure of ROTSE. It is
important to note that the overlap between the internal shocks signal
(the GRB) and the early afterglow (the optical and the X-ray) was
expected.  In the internal shocks scenario long bursts
are produced by thick shells, which, in turn,  are produced by a central 
engine operating for a long time.
Sari (1997) have shown that for this case, there should be an overlap 
between the internal shock emission and the external shock emission,
in agreement with the observations.

\section{The Reverse Shock Evolution}

An exact calculation of the reverse shock evolution requires a
detailed understanding of the magneto-hydrodynamics of relativistic
collisionless fluids and their behavior behind strong shocks. However,
surprisingly good qualitative picture can be obtained by treating the
matter as a fluid, using the simplest assumptions (equipartition and
random orientation) on the magnetic field evolution.

After the reverse shock has passed through the ejecta, the ejecta
cools adiabatically. We assume that it follows now the Blandford McKee
(1976) self-similar solution (recall that strictly speaking this
solution deals only with the ISM material).  In this solution a
given fluid  element evolves with a bulk Lorentz factor of $\gamma
\sim R^{-7/2}$. Since the observed time is given by $T \sim
R/\gamma^2c$ we obtain
\begin{equation}
\label{scale_gamma}
\gamma \sim T^{-7/16}.
\end{equation}
Similarly, the internal energy density evolves as $e \sim R^{-26/3}
\sim T^{-13/12}$, the particle density evolves as $n \sim R^{-13/2}
\sim T^{-13/16}$ and therefore the energy per particle, or the particle
Lorentz factor behaves like
\begin{equation}
\label{scale_gamma_e}
\gamma_e \sim T^{-13/48}.
\end{equation}

The simplest assumption regarding the magnetic field is that its
energy density remains a constant fraction of the internal energy
density. In this case we obtain $B \sim \sqrt{e} \sim T^{-13/24}$.
Other evolution of the magnetic field are possible if the
magnetic field has a defined orientation (Granot, Piran \& Sari 1998).

We assume that the reverse shock has accelerated the electrons to a
power-law distribution. However, once the reverse shock crossed the
ejecta shell, no new electrons are accelerated. All the electrons
above a certain energy cool, and if the cooling frequency, $\nu_c$, 
is above the
typical frequency, we are left with a power law electron distribution
over a finite range of energies and Lorentz factors. Each electron now
cools only due to the adiabatic expansion with its Lorentz factor
proportional to $T^{-13/48}$.

Once the reverse shock has crossed the ejecta shell the emission
frequency drops quickly with time according to $\nu_e \sim \gamma
\gamma_e^2 B \sim T^{-73/48}$. Given that the total number of
radiating electrons $N_e$ is fixed the flux at this frequency falls
like $F_{\nu_e} \sim N_e B \gamma \sim T^{-47/48}$.  Below the typical
emission frequency, $\nu_m$, we have the usual synchrotron low energy
tail and for these low frequencies the flux decreases as $F_\nu \sim
F_{\nu_m} (\nu/\nu_m)^{1/3} \sim T^{-17/36}$.

Above $\nu_m$ (and below $\nu_c$) the flux falls sharply as $F_\nu
\sim F_{\nu_m} (\nu/\nu_m)^{-(p-1)/2}$. For $p=2.5$ this is
about $F_\nu \sim T^{-2.1}$. Both $\nu_m$ and $\nu_c$ drop as
$T^{-73/48}$, since all electrons cool by adiabatic expansion only. 
Once $\nu_c$ drops below the observed frequency
the flux practically vanishes (drops exponentially with time).

\section{The Reverse Shock Emission and GRB990123  
Observations}

The initial decay of the optical flux after the second ROTSE exposure
is $T^{-2}$.  In agreement with the crude theory predicting $-2.1$ .
This means that $\nu_m$, the typical synchrotron frequency of the
reverse shock, is below the optical bands quite early on.  Using the
estimates of the peak value of reverse shock $\nu_m$ from SP99 we
obtain
\begin{equation}
\nu _{m}=1.2\times 10^{14}\left( \frac{\epsilon _{e}}{0.1}\right)
^{2}\left( \frac{\epsilon _{B}}{0.1}\right) ^{1/2}(\frac{\gamma _{0}}{300}
)^{2}n_{1}^{1/2} \le 5 \times 10^{14}.
\end{equation}
This shows that the initial Lorentz factor of this burst was not too
high.  Using the equipartition values $\epsilon_e \sim 0.6$ and
$\epsilon_B \sim 0.01$ and $n_{1}=5$ inferred for GRB970508 
(Wijers \& Galama, 1998, Granot, Piran \& Sari, 1998b)
we find that the initial Lorentz factor was rather modest:
\begin{equation}
\label{gamma_0}
\gamma_0 \sim 200 .
\end{equation}
This is in agreement with the lower limit estimates, based on the pair
creation opacity (Fenimore, Epstein, \& Ho, 1993; Woods \& Loeb, 1995;
Piran, 1995; Baring \& Harding, 1997).

The Lorentz factor at the beginning of the self similar
deceleration, i.e., at the time of the afterglow peak $\sim 50$s, 
$\gamma_A \sim 220$ is independent of the initial Lorentz factor
of the flow (SP99). This is very close to our estimated 
initial Lorentz factor. It shows that the reverse
shock is only mildly relativistic and the initial   Lorentz factor of
its accelerated electrons' random motion is:
\begin{equation}
\label{gamma_e_0}
\gamma_m \sim 630 \epsilon_e .
\end{equation}

Emission from the reverse shock can also explain the radio
observations: a single detection of $\sim 260\mu$Jy one day after the
burst.  If the reverse shock emission peaked in the optical at $\sim
50$sec and the peak frequency decayed in time as $T^{-73/48}$ then the
peak frequency should have reached $8.4$GHz after $\sim
19$hours. Scaling the observed optical flux of $1$Jy, as $T^{-47/48}$
to $19$ hours the expected flux at $\nu_m=8.4$GHz is $840\mu$Jy. From
that time on the 8.4GHz flux decays as $T^{-2.1}$. The emitted 8.4GHz 
flux is therefore given by
\begin{equation}
\label{radioemitted}
F_\nu=
   \cases{840 \mu {\rm Jy} (T/19h)^{-2.1}   & $T>19$ hours \cr
          840 \mu {\rm Jy} (T/19h)^{-17/36} & $T<19$ hours },
\end{equation}
so that at $1.2$ days,
when radio was detected, we expect a flux of $350\mu$Jy, amazingly
close to the observations. Equation \ref{radioemitted} is also compatible
with all later upper limits, see figure 2.

\begin{figure}[tbp]
\begin{center}
\epsscale{1.} \plotone{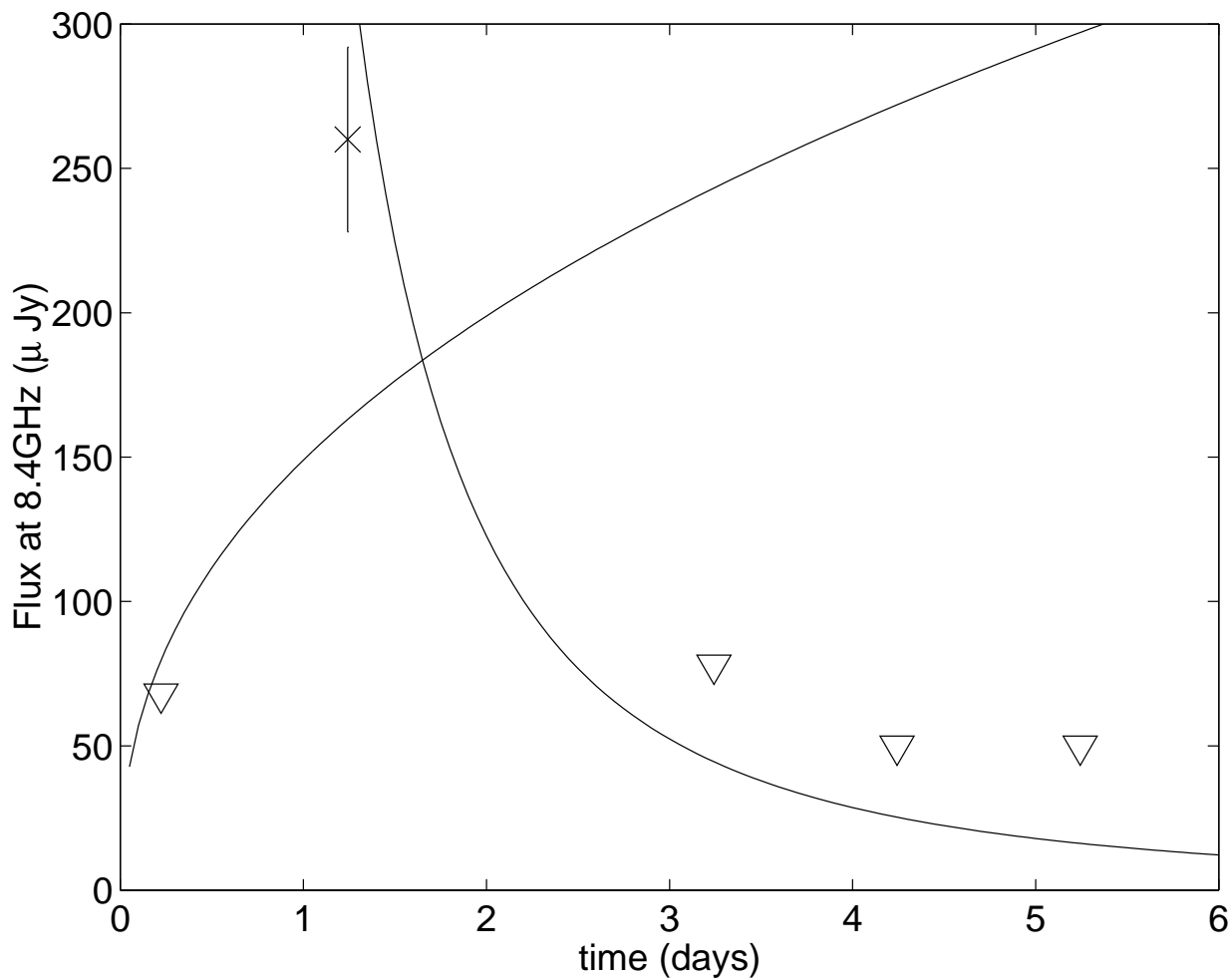}
\end{center}
\caption{The single radio detection at 8.4GHz (x) and the upper limits
measured during the first 5 days (triangles). 
The decaying solid line gives the
emission from the reverse shock ignoring self absorption. The rising
solid line given the maximal flux allowed by black body emission. The
expected emission is the minimum between the two. While overall the
fit is reasonable, the flux seems to rise faster than the theory predicts 
between the 6 hours
after the burst and 1.2 days.  This might be either the result of
moderate scintillation or moderate absorption from the forward shock
that is more significant in the early epoch.  }
\label{fig2}
\end{figure}

Equation \ref{radioemitted} yields a 8.4GHz flux of $1.5$mJy  after six
hours, which is way above the upper limit of $64\mu$Jy.
However, strong self absorption of the reverse shock radio
emission took place at this stage and this suppressed this emission. 
When accounting for that, the
resulting emission would be the minimum between the estimate of equation
\ref{radioemitted} (ignoring self absorption) and the black body upper
limit.
This upper limit of  black body emission 
from the reverse shock can be estimated by (Katz \& Piran, 1997, SP99):
\begin{equation}
\label{BB}
F_{\nu,BB}=\frac {2 \nu^2} {c^2} \pi \gamma \gamma_e m_e c^2 (R_\perp/D)^2=
150\mu{\rm J} (T/1{\rm day})^{5/12}.
\end{equation}
Note that while the  emission
estimates used only scaling with time of the observed early optical
flux, the  black body upper limit is  more model dependent and
possibly less reliable.
The scaling in the last expression as well as the numerical
coefficient use the scaling of $\gamma$ and $\gamma_e$ with time (equation 
\ref{scale_gamma} and \ref{scale_gamma_e}) together with their inferred 
initial value from the initial afterglow time and peak (equations
\ref{gamma_0} and \ref{gamma_e_0}).  We used $R_\perp \sim 4.6 \gamma c T$
where the numerical coefficient is appropriate for a fast decelerating
shell (see Sari 1997,1998; Waxman 1997a; Panaitescu \& M\'esz\'aros
1998), and the relevant distance is $D=D_L/\sqrt{1+z}\sim 1.7 \times
10^{28}$cm for a $\Omega=1$, $h=65$ universe and assuming $z=1.61$.
Shown, on Fig. 2 is also the upper limit to the radio emission 
according to a black body spectrum.  This upper
limit from black body emission also accounts for the lack of 4.88GHz
reported by Galama et al. (GCN212).

We now turn to estimate the initial (50sec) cooling frequency $\nu_c$.
Note that initially, this frequency is the same for the forward and the 
reverse shock (SP99). A simple estimate can
be obtained from the the temporal slope of the late afterglow (forward shock) 
light curve and its spectrum, which
are compatible with the predicted spectrum
(Sari, Piran \& Narayan, 1998) of 
slow cooling electrons (Bloom et at., GCN240). This
indicates that the forward shock cooling frequency is $\nu_{c,f}(2{\rm
days}) \ge 4 \times 10^{14}$Hz leading to $\nu_c(50{\rm
s})=\nu_{c,r}=\nu_{c,f} \ge 2.5 \times 10^{16}$Hz. Detection of a break
in the optical flux later on can be used to replace this inequality by
a more solid number. 

A similar lower limit can be put using the fact that the reverse shock
was seen for more than 650sec. The reverse shock cooling frequency at that time
is therefore higher than $5 \times 10^{14}$Hz. Scaling it back to 50sec according
to $T^{-73/48}$ we get $\nu_c(50{\rm s}) \ge 2 \times 10^{16}$Hz.

A more speculative constraint on $\nu_c$ can be obtained from the GRB
spectrum itself (SP99). The decreasing GRB spectrum below a few
hundred keV implies that $h \nu_c \ge 100 $keV.  Otherwise the low
energy flux would have increased at low frequencies like
$\nu^{-1/2}$. This constraints $\nu_c \ge 2 \times 10^{19}$Hz. This
holds for the site producing the GRB (internal shocks in our model).
However, the observed $\gamma$-ray emission during the end of the GRB
is probably dominated by the forward shock, as suggested by the
smoother temporal profile and by the softer spectrum. This means that
a similar constraint applies to the initial $\nu_c$ of the forward shock, 
which is the same as the reverse shock one.  
If this rough estimate of $10^{19}$Hz is correct then this break in the 
late optical light curve is expected to be only after about 40 days.

The reverse shock model can  also be confronted with the observed
optical to $\gamma$-ray energy ratio. Using the table in SP99 and
the estimated synchrotron frequency at $\sim 50$sec $\nu_m \sim 4
\times 10^{14}$Hz and the cooling frequency $\nu_c \sim 10^{19}$Hz we
find the optical fluence to be $4\times 10^{-3}$ of the GRB fluence. 
This is about a factor of five higher than the observed
fraction, a reasonable agreement, considering the crudeness of the
model. Note that this model assumes that the reverse shock
contains the same amount of energy as the whole system. This can be
of course lower by a factor of a few.

\section{The Forward Shock}
The forward shock that propagates into the ISM is considered by now
as the classical source of the afterglow (Katz, 1994; M\'esz\'aros \& Rees
1997; Vietri, 1997; Wijers et al., 1997; Waxman, 1997b; Katz \& Piran,
1997). After a possible short radiative phase it becomes
adiabatic and it acquires the Blandford McKee profile. It then
expands self-similarly until it becomes non-relativistic.  In GRB990123 it has
produced some of the prompt soft $\gamma$-rays and X-rays observed
late during the burst by BATSE, and the WFC. It has continued to
produce the X-ray and the late optical and IR emission.

The initial decline of the X-ray suggests that already initially the
typical synchrotron frequency was below the 1.5-10keV band.  The late
slope of the light curve of the optical afterglow agrees well with
other power law decays seen in other afterglows. This suggests that we
see an adiabatic decay phase. There could have been an early radiative
phase, but if there was one it was shorter than the 3.75 hours gap
before the first optical observations. This is in agreement with
expectations (Waxman 1997, Granot et al., 1998a).  The decline from
$3.75$ hours onwards in the R band suggests that already at this stage
the typical synchrotron frequency, $\nu_m$ was below this
band. Extrapolating this back to 50sec we get $\nu_m \le 2 \times
10^{18}$Hz, consistent with the above discussion.  Moreover, the ratio
between $\nu_m$ of the forward and the reverse shock should be
approximately $\gamma^2$. Using the two estimated values in the
initial time we find that $\gamma \sim 70$. This is a factor of 3
lower than completely independent estimate in equation
\ref{gamma_0}. Again we consider this as a rather good agreement in
view of the crudeness of both estimates.  A short initial radiative
phase could even improve this agreement.

The observed temporal decay slope of the X-ray ($-1.35$ and $-1.1$) and
optical ($-1.13$) from the forward shock are comparable. An X-ray slope
steeper by a $1/4$ is predicted (Sari, Piran and Narayan 1998) 
if the cooling frequency is between the X-rays and the optical, which
seems to be the case in this burst.
With future data and a careful analysis this prediction could be tested.

\section{Discussion}

The discovery of prompt optical emission during a GRB have opened a
new window to explore this remarkable phenomenon. The lack of
correlation between the optical and $\gamma$-ray emission is a clear
indication that two different processes produce the emission in those
two different bands. The emission from these two processes reach the
observers simultaneously. These findings are in a perfect agreement with the
internal-external model. Fenimore, Ramirez-Ruiz and Wu (1999) reach
into the same conclusion on the grounds of the burst's temporal structure.

The strong prompt optical emission was predicted (SP99) to arise from
the reverse external shock. We see here that various features of this
emission, in particular the overall fluence and the decay slope agree
well with the predictions. This reverse shock also explains
the origin of the transient radio observation a day after the burst.

The light curves of the X-ray and the late optical afterglow agree
well with the, by now, ``classical'' afterglow model. According to
this model this emission is produced by the forward external shock.
We expect a somewhat different slope for the X-ray and optical light
curves. However at present the data is not good enough to test this
prediction. It remains to be seen if this could be tested in the
future.  We also expect that radio emission would show up on a time
scale of weeks. The source of this emission would also be the forward
external shock.

Already now we were able to determine some of the parameters of
GRB990123. Specifically we were able to estimate the initial Lorentz
factor and the Lorentz factor three days after the burst.  Future
radio and optical observations will enable us to determine the rest of
the parameters of GRB990123, allowing a refinement of these
calculations and further tests of the theory.

The late optical light curve is well fit by a single power law without
any break.  The index of this slope is approximately the one predicted
by the spherical afterglow model. These facts suggest that so far
there was no transition from a spherical like  to an expanding jet
behavior.  Such a transition is expected, for a relativistic jet, when
the Lorentz factor reaches the value $\theta^{-1}$, where $\theta$ is
the opening angle of the jet (Rhoads, 1997). Such a transition would
lead to a break in the light curve and to a decrease in its index by
one.  
Since the theory gives a Lorentz factor of about six at seven days, 
these observations set a lower limit on the beaming angle of
GRB990123 to be $\theta \ge 0.15$. The energy budget could still be
``rescued'' if a break is seen soon. Otherwise, this indicates that
GRB990123 is as powerful as the isotropic estimates suggest!

The coincidence of nearby galaxy the strength of the burst have led to
the speculation that GRB990123 has been magnified by a gravitational
lens (Djorgovski et al., GCN216). There have been some suggestions
that this is unlikely. We stress that our analysis (apart from the
black body emission in equation \ref{BB}) is independent of whether the 
event was lensed or not and independent of its redshift.

\acknowledgments
This research was supported by the US-Israel BSF grant 95-328, by a
grant from the Israeli Space Agency and by NASA grant NAG5-3516.
Re'em Sari thanks the Sherman Fairchild Foundation for support, 
and Eric Blackman and David Band for discussions and useful remarks.
Tsvi Piran thanks Columbia University and Marc Kamionkowski for 
hospitality while this research was done and Pawan Kumar for 
many helpful discussions.

\end{document}